\documentstyle[prl,aps,floats,psfig]{revtex}
\tighten
\begin{document}

\title{Nonminimal Global Monopoles and Bound Orbits\\ } 

\author{Ulises Nucamendi\footnote{e-mail: ulises@fis.cinvestav.mx} \\
\small {\it Departamento de F\'{\i}sica}\\
\small {\it Centro de Investigaci\'on y de Estudios Avanzados del I.P.N.}\\
\small {\it A. P. 14-741, M\'exico, D. F. 07000,  M\'exico}\\
\small {\it and Instituto de Ciencias Nucleares} \\ 
\small {\it Universidad Nacional Aut\'onoma de M\'exico} \\ 
\small {\it Apdo. Postal 70-543 M\'exico 04510 D.F, M\'exico}\\  
\vspace*{0.5cm}
Marcelo Salgado\footnote{e-mail: marcelo@nuclecu.unam.mx} and 
Daniel Sudarsky\footnote{e-mail: sudarsky@nuclecu.unam.mx} \\ 
\small {\it Instituto de Ciencias Nucleares} \\ 
\small {\it Universidad Nacional Aut\'onoma de M\'exico} \\ 
\small {\it Apdo. Postal 70-543 M\'exico 04510 D.F, M\'exico}. }
\maketitle

\centerline {\bf Abstract}
We perform a numerical analysis of the gravitational field of 
a global monopole coupled nonminimally to gravity, and find 
that, for some given nonminimal couplings (in contrast with the minimal 
coupling case), there is an attractive 
region where bound orbits exist. 
We exhibit the behavior of the frequency shifts that would be 
associated with ``rotation curves'' of stars in circular orbits 
in the spacetimes of such global monopoles.
\vskip 1.5cm
\noindent PACS number(s): 11.27.+d, 04.40.-b, 98.62.Gq

\bigskip
An interesting property of global monopoles is that they are
configurations with energy density decreasing with the distance as
$r^{-2}$ \cite{vilenkin}. 
This is very suggestive in view of the fact that, naively
at least, this is precisely what seems to be required to provide a 
natural explanation for the flatness of the rotation curves in
spiral galaxies. Moreover by assuming the existence of a global
monopole in a typical galaxy the total Newtonian
mass contribution of the portion of the global monopole contained
within $r_{gal}$ is found to be 
$M \approx \alpha r_{gal}/2$, 
where $\alpha = 8 \pi G \eta^2$ is the deficit angle of the 
global monopole spacetime. 
If we take the radial extent of a galaxy to be 
$r_{gal} \approx 15$\,Kpc and 
consider a typical grand unified theory with a symmetry 
breaking scale $\eta \approx 10^{16}\,$ GeV, this mass turns out to be 
$\approx 10^{69} \,$ GeV which is ten times the total mass due to the 
$10^{11}$ solar-mass stars in the typical galaxy: 
$M_{stars} \approx 10^{68}\,$ GeV. 
This again seems to be 
precisely what is called for in the observations \cite{wald}.
Finally, and what seems to be an added bonus, if we assume that 
the field of the monopole extends on average a distance of
10 galactic radii from the galaxy, where the
configuration presumably coincides with that of the monopole
centered in the neighboring galaxy, the total mass associated
with each monopole (and thus with each galaxy) turns out to be 
100 times that of the galaxy. Thus the total contribution
of the monopoles to the average density in the universe is of
the right order of magnitude to account for the inflationary 
prediction of a universe with critical density. This is in fact
the basic argument that allows one to place bounds on the number
density of global monopoles present in the universe \cite{hiscock},
the exact nature of which depends of course on the value taken 
for $\eta$. Indeed global monopoles
have been shown to be able to trigger inflation on their own for suitably
high $\eta$ \cite{vilenkin2}, \cite{sakai}, \cite{ion}. \\

\begin{figure*}[h]
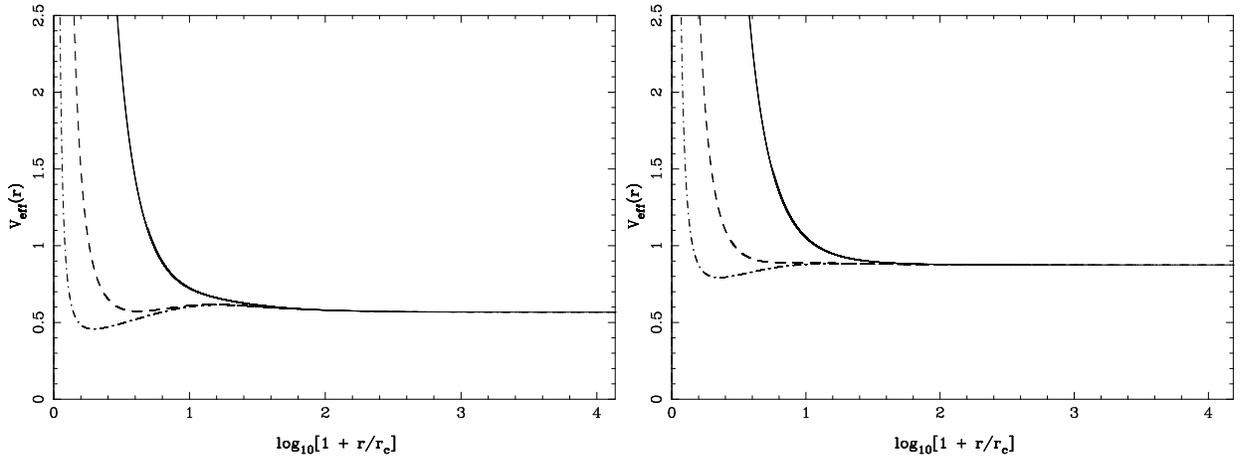

\centerline{
\psfig{file= poteff-2.ps,angle=-90,width=8.1cm} 
\psfig{file= poteff-21.ps,angle=-90,width=8.1cm}}
\vspace*{0.5cm}
\caption[]{\label{f:velo1}
Functional dependence of the effective potential 
$V_{eff}$ vs $\tilde r$ for the case
$\xi=-2$, $\alpha=0.43$ (left fig.) 
and $\alpha=0.125$ (right fig.). 
For each configuration we show three values of the 
angular momentum: $L=4$ (solid line),
$L=1$ (dashed line) and $L=0.3$ (dash-dotted line).
Here $r_c \equiv (\eta\lambda^{1/2})^{-1}$.}  
\end{figure*}

Unfortunately this picture doesn't hold up to scrutiny at the next level
of analysis because of two problems: 1) upon substitution 
of Newtonian gravity by general relativity, it turns out \cite{vilenkin},
\cite{harari}
that the linearly divergent mass has, at large distances, an effect
analogous to that of a deficit solid angle plus that of a tiny mass
at the origin. 
The mass of this ``monopole core'' is about
$0.8 \,\alpha$. In fact Harari and Loust\'o \cite{harari} showed
that this small core mass is negative and produces a repulsive 
potential. They studied the motion of test particles, in the large 
distance region, in the spacetime of a global monopole, concluding 
that there are no bound orbits; 2) the fact that 
the monopole configuration is rather unique, 
in the sense that it is basically independent of
the ordinary matter content in the corresponding galaxy, 
which conflicts with the fact that there is 
a rather large range of galactic masses for which the dark matter
component is about 10 times more massive than the ordinary matter 
component \cite{wald}. \\
\\
In spite of these remarks the initially mentioned results are in our
view still very suggestive, specially if we look at the remarkable
degree of universality that seems to emerge  from the systematic
study of galactic dynamics \cite{persic}: 1) the ordinary matter 
content is such a good indicator of the mass
of the dark matter component; 2) the universality of the form of the
radial distribution of the dark matter content; 3) other 
seemingly coincidental facts as scaling properties between dark and
luminous galactic structure parameters. 
We, therefore, believe that the general idea of
monopoles as seeds of galaxy formation, and models for the galactic
(and perhaps cosmological) dark matter deserve to be further explored.\\
\\
The purpose of this Letter is to illustrate the possibilities in
these models by 
considering the simplest nontrivial modification of the global 
monopole picture: nonminimally coupled global monopoles. 
In this work we will show that the situation regarding the problems
alluded to above, can change dramatically with the introduction of a
nonminimal coupling. \\
\\
Let's consider the simplest such possibility, namely a theory of 
a triplet of scalar fields 
$\phi^a$, $a=1,2,3$, nonminimally coupled (NMC) to gravity with 
global O(3) symmetry which is broken spontaneously to U(1).
The Lagrangian for the model is
\\
\begin{equation}
{\cal L} = \sqrt{-g} \left({ 1\over 16\pi } R + 
F(R, \phi^a\phi^a) \right) -
\sqrt{-g} \left[ {1\over 2}(\nabla \phi^a)^2 
+ V(\phi^a\phi^a) \right] \,\,\,\,.
\label{lag}
\end{equation}
\\
where $V(\phi^a\phi^a)$ is a potential that results in the breaking
of the symmetry and is usually taken to be the Mexican hat potential 
$V(\phi^a\phi^a)={\lambda\over 4}(\phi^a\phi^a-\eta^2)^2$, and $F$ 
is an arbitrary function of two variables. Note that in these 
theories the resulting potential for the scalar fields (the terms
in the Lagrangian that are independent of the scalar field gradients)
is 
\\
\begin{eqnarray}
V(\phi^a\phi^a)_{res} = V(\phi^a\phi^a) - F(R, \phi^a\phi^a) \,\,\,\,.
\end{eqnarray}
\\
Therefore, the ordinary matter would affect the location of the
minimum of the effective potential through its effect on $R$, thus
avoiding the scenario where the monopole configuration is universal. 
This opens the possibility to recover the correlation between the masses 
in the dark and ordinary matter components of the galaxy that were
alluded before. \\
\\
Next we focus on the existence or lack of existence of bound orbits
in the corresponding spacetimes. To do this we will look at a specific
example corresponding to the choice $F = (\xi \phi^a\phi^a) R$
(where $\xi$ is a constant) and proceed to construct the resulting
spacetime and analyze the motion of test particles. 
We will specialize to the case of a spherical and static solution 
to the coupled field equations. The configuration describing a 
monopole is taken as usual
\\
\begin{eqnarray}
\label{conf}  
\phi^a &=& \eta f(r) \frac{x^a}{r},
\end{eqnarray}
\\
with $x^a x^a = r^2$, so that we will actually have a monopole 
solution if $f \rightarrow 1$ at spatial infinity. 
We adopt the following metric:
\\
\begin{equation}
ds^2 = -N^2(r) dt^2 + A^2(r)dr^2 + r^2d\theta^2
+ r^2\sin^2\theta d\varphi^2 \,\,\,\,\,,
\label{RGMS}
\end{equation}
\\
and analyze solutions of the gravitational and
scalar fields equations describing global monopole
configurations and the resulting space-time. Owing to the 
complexity of the resulting equations, we will
perform a numerical analysis. To do so we adopt the 
following variables:
\\
\begin{eqnarray}
\nu (\tilde r) &=& {\rm ln}[N(\tilde r)] \,\,\,\,\,,\\
\label{tildenu}
A(\tilde r) &=& 
\left(1- \alpha -
\frac{2 m(\tilde r)}{\tilde r}\right)^{-1/2}
\,\,\,\,\,,
\label{AA}
\end{eqnarray}
\\
where
\\
\begin{equation}
\label{alfa}
\alpha= \frac{\Delta}{1+2\xi\Delta}, \,\,\,\,\,\,
\Delta= 8\pi \eta^2 \,\,\,,
\end{equation}
\\
where we have introduced the dimensionless quantity
$\tilde r \equiv r/r_c \equiv \eta\lambda^{1/2} \cdot r$.  \\
\\
The solutions are obtained by the standard shooting method
\cite{recipes} with boundary conditions ensuring a regular
origin, i.e., $f(0)=0$, $m(0)=0$,
$m(0)_{,\tilde r}= -\alpha/2$, and that the spacetime is 
asymptotically flat but for a deficit angle \cite{ulises}, i.e., 
$f(\infty)=1$, $\nu(\infty)=\frac{1}{2}\ln\left[1 - \alpha\right]$.
The ``shooting parameter'' is taken to be $f(0)_{,\tilde r}$.
\\  
Then we calculate the ADM (gravitational) mass of these configurations, 
(see \cite{ulises} for a rigorous definition of the ADM mass for 
spacetimes that are asymptotically flat but for a deficit angle)
which can be easily evaluated from the integral 
\\
\begin{equation}
M_{\rm ADM\alpha} = {\rm lim}_{\tilde r\rightarrow\infty} \,\,
m (\tilde r) = \int_{0}^{\infty} 
\left(4\pi \tilde r^2 T^{t}_{t}(\tilde r) - 
\frac{\alpha}{2}\right) d\tilde r \,\,\,\,.
\end{equation}
\\        
For all values of $\xi$ that we analyzed the resulting ADM mass 
was negative as in the minimal coupling case (here we must remind 
the reader that the ADM mass for quasi-asymptotically flat spacetimes 
does not share the positiveness property of the ADM mass of 
asymptotically flat spacetimes). 
For $\xi=-2$, 
for example (corresponding to the configuration we will 
analyze in more detail) the resulting ADM mass is 
$\approx -2.5 \,\alpha$.  \\
\\
We now study the motion of test particles around a 
global monopole.
To do this we consider the ``equatorial geodesics'' for the metric
(\ref{RGMS}) (i.e., geodesics lying on the plane 
$\theta = \pi/2$). We denote the 4-velocity of a test particle as
$u^{\mu} = (\dot{t}, \dot{r}, 0, \dot{\varphi})$, 
where the dot stands for derivation with respect to an affine 
parameter which in the case of a massive particle 
can be taken as the proper time.
The resulting equation is 
\\
\begin{equation}
\label{egrpunto}
\frac{\exp [2\nu (\tilde r)]}
{\left[1- \alpha -\frac{2 \tilde m(\tilde r)}{\tilde r}\right]} 
(\dot{\tilde r})^2 + \exp [2\nu (\tilde r)] 
\left[ \frac{L^2}{(\tilde r)^2} + 1 \right] = E^2  \,\, ,
\end{equation}
\\
where $E$ and $L$ are the energy and the angular momentum 
per unit of mass of the test particle respectively. 
This equation
shows that the radial motion is the same as that 
of a particle with a ``position dependent mass'' with energy $E^2/2$ 
in ordinary one-dimensional, nonrelativistic mechanics moving 
in the effective potential
\\
\begin{equation}
V_{eff} = \exp [2\nu (\tilde r)] 
\left[ \frac{L^2}{(\tilde r)^2} + 1 \right].
\label{veffe}
\end{equation}
\\
In Figure \ref{f:velo1} we show the effective potential 
$V_{eff}$ (see eq. (\ref{veffe}))
for the deficit angles $\alpha=0.43$ and $\alpha=0.125$ 
respectively and for three values of the angular momentum; 
we note the presence of a potential well and 
therefore the existence of stable circular orbits. \\
\\

\begin{figure*}[h]
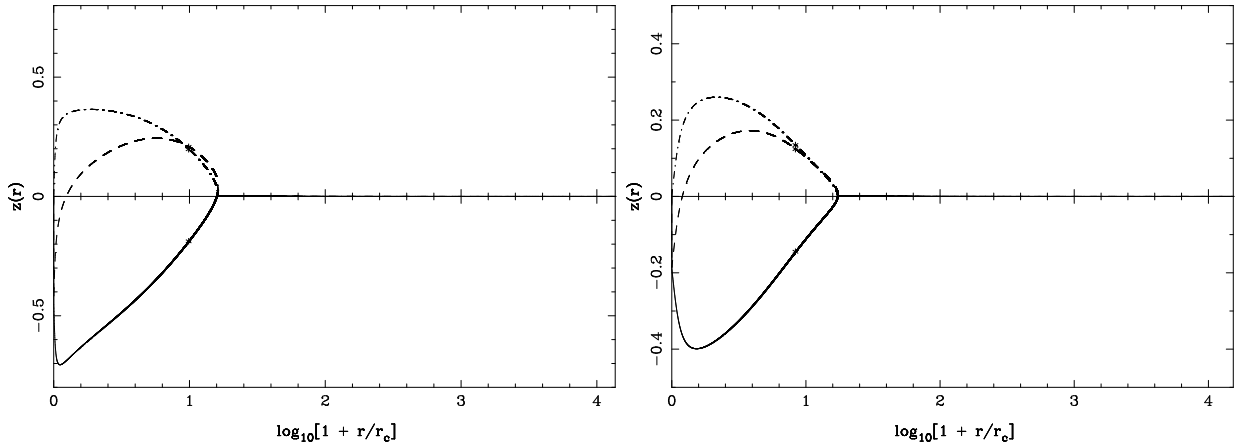

\centerline{
\psfig{file= rs.ps,angle=-90,width=8.1cm} 
\psfig{file= rs1.ps,angle=-90,width=8.1cm}}
\vspace*{0.5cm}
\caption[]{\label{f:rs}
Functional dependence of $z_+$ (dashed line),
$z_-$ (solid line) and $z_D$ (dash-dotted line)
vs $\tilde r$ for circular orbits in the case
$\xi=-2$, $\alpha=0.43$ (left fig.) and $\alpha=0.125$
(right fig.). The asterisk depicts the location of the 
radius beyond which the stable circular orbits cease to exist
($r \sim 9 \,r_c$).}  
\end{figure*}

The rotation curves (RC) of spiral galaxies
are deduced from the red and blue shifts of the 
emitted radiation by stars moving in ``circular orbits'' 
on both sides of the central region \cite{rubin1}. 
We therefore evaluate the frequency shift for a light
signal emitted from a ``test star'' in circular orbit 
in the spacetime of a global monopole and received by a 
static observer at infinity.
The two maximum and minimum frequency shifts 
(associated with a receding or an approaching star respectively) 
turn out to be: 
\\
\begin{equation}
z_{\pm} = 
1 - \left[ \frac{(1-\alpha)}{ N(\tilde r) ( N(\tilde r) - 
\tilde r N(\tilde r)_{,\tilde r} )} \right]^{1/2} 
\left[ 1 \mp 
\left( \frac{ \tilde r N(\tilde r)_{,\tilde r} }{ N } \right)^{1/2} \right]  
\,\,,
\label{redshift1}
\end{equation}
\\
with $z$ defined as usual,
\begin{equation}
z \equiv \frac{\Delta \omega}{\omega} = 
\frac{\omega_R - \omega_E}{\omega_R}  \,\,,
\label{redshift2}
\end{equation}
\\
where $\omega_E$ ($\omega_R$) is the emitted (detected)
frequency. We can define $z_D = \frac{1}{2} (z_+ - z_-)$ 
which would be the quantity that is operationally 
identified with the velocity $|v(\tilde r)|$ of the RC 
of spiral galaxies
\\
\begin{equation}
z_{D} = 
\left[ \frac{(1-\alpha)}{ N(\tilde r) ( N(\tilde r)- 
\tilde r N(\tilde r)_{,\tilde r} )} \right]^{1/2} 
\left[ \frac{ \tilde r N(\tilde r)_{,\tilde r} }{ N } \right]^{1/2}   \,\,.
\label{rsm}
\end{equation}
\\
In Figure \ref{f:rs} we show $z_+$ (dashed line),
$z_-$ (solid line) and $z_D$ (dash-dotted line) 
as functions of $\tilde r$ for the cases $\alpha = 0.43$
(left fig.) and $\alpha = 0.125$ (right fig.) 
respectively. We note that even for this very simple model
the figures that would correspond to the rotation curves contain
a relatively ``flat region'' within the values of $r$ corresponding
to the stable orbits (i.e. the behavior of $z_D$ near its maximum).  \\
\\

We are of course not suggesting that this simple model 
is in any way a realistic candidate for a natural 
explanation of the phenomena associated with the RC of spiral 
galaxies. However, the fact that the two main 
objections against the study of global monopoles in 
this context can be removed by such a
simple modification as the introduction of a nonminimal coupling and 
the surprising coincidences mentioned 
at the beginning of this letter, provide very suggestive motivations for  
pursuing in the analysis of such possibilities. We must, nevertheless, 
emphasize that monopole-antimonopole pairs annihilate
rather efficiently \cite{rhie} and therefore
the viability of this scenario,
as applied to the galactic rotation curves,
will also depend strongly on the degree of asymmetry in the
number of monopoles and antimonopoles in the early universe.\\

In conclusion we have shown that the qualitative difficulties of 
the original monopole scenario for explaining the rotation curves 
of galaxies can be solved by the introduction of nonminimal 
coupling. The challenge for the future is to look for a specific 
realization of this more complicated scenario that combines the 
qualitative success of the latter model with the quantitative 
coincidences of the former.  

\acknowledgments
We wish to thank R. M. Wald for helpful
discussions and also to J. Guven for his comments. 
M.S. and D.S. would like to acknowledge partial 
support from DGAPA-UNAM project IN121298 and to thank the
supercomputing department of DGSCA-UNAM.

\end{document}